# Observation of the spin-based plasmonic effect in nanoscale structures

Y. Gorodetski, A. Niv, V. Kleiner and E. Hasman[*]

*Micro and Nanooptics Laboratory, Russel Berrie Nanotechnology Institute,*

*Technion - Israel Institute of Technology, Haifa 32000, Israel*

[*]**e-mail: mehasman@tx.technion.ac.il**

**Abstract**

Observation of surface-plasmon phenomena that are dependent upon the handedness of the circularly polarized incident light (spin) is presented. The polarization-dependent near-field intensity distribution obtained in our experiment is attributed to the presence of a geometric phase arising from the interaction of light with an anisotropic and inhomogeneous nanoscale structure. A near-field vortex surface mode with a spin-dependent topological charge was obtained in a plasmonic microcavity. The remarkable phenomenon of polarization-sensitive focusing in a plasmonic structure was also demonstrated.

The interaction of light with metallic subwavelength structures exhibits various anomalous effects such as extraordinary optical transmission [1] and beaming [2]. These effects have been elegantly explained by a mechanism involving the coupling of light to collective surface-confined electronic oscillations known as surface plasmon-polaritons (SPPs). Extensive research has been carried out in the field of electromagnetic surface waves due to its technological potential and fundamental implications. Additional exciting phenomena were studied such as unidirectional plasmon coupling [3], plasmon focusing [4,5], waveguiding and interferometry [6,7], enhanced coherent thermal emission [8,9], surface-enhanced Raman scattering [10]), planar optical chirality [11,12] and superresolution [13,14]. The handedness of the light's polarization (optical spin up/down) provides an additional degree of freedom in nanoscale photonics. Here, we observe a spin-dependent behaviour of SPPs resulting from a geometric Berry phase that was experimentally found in the interaction of light with metallic anisotropic and inhomogeneous nanoscale structures. We demonstrate surface structures for which the SPP near-field intensity distribution depends upon the polarization state of the incident beam. We obtain a plasmonic vortex with a spin-dependent geometric phase in microcavities with anisotropic inhomogeneous boundaries. The remarkable phenomenon of a spin-dependent shift of the focal spot in a focusing plasmonic device is also measured. We believe that the interaction between the spin of the photons and the optical medium can be exploited to enable the development of a new generation of nano-devices.

The proposed anisotropic inhomogeneous plasmonic structure is produced on top of a thin metal film evaporated onto a glass plate (Fig. 1(a,b)). The element consists of a spiral Bragg grating with a central defect, surrounded by a coupling grating, both of which were etched to a depth of 50nm by a focused ion beam (FEI Strata 400s dual beam system, $Ga^+$, 30keV, 46pA). The thickness of the metal (Au) was chosen to be 100nm in order to prevent any direct transmission of light. The grating was designed to couple normally incident light at $\lambda_0 = 532$nm to a surface-plasmon wave, while the Bragg grating with a central defect serves as a plasmonic microcavity for the surface wave. Therefore, the actual coupler period and the Bragg period were chosen to be 0.5μm and 0.25μm, respectively. The outer diameter of the structure was 10μm and the small radius of the spiral cavity was chosen to be $r_0 = 1.32$μm. These parameters are also consistent with the surface-plasmon coherence length [16] $L_c \propto 1/\text{Im}(k_p)$, which is about 3μm for gold at this wavelength.

The structure was illuminated by circularly polarized light which is denoted henceforth with a spin state $|\sigma_\pm\rangle$, where $\sigma_+ = 1$ stands for right-handed circularly polarized light and $\sigma_- = -1$ for left-handed circularly polarized light. As the illumination source, we used a solid state laser at $\lambda_0 = 532$nm ("Verdi", Coherent). The intensity in the near-field of the plasmonic cavity was measured by a Near-field Scanning Optical Microscope (NSOM; Nanonics Multiview 2000) in a non-contact mode (see Fig 1(a)). The captured intensity distribution is presented in Fig. 1(c,d). The near-field intensity distribution was also calculated by a Finite Difference Time Domain (FDTD) algorithm as provided in Fig. 1(e,f). Surprisingly, the measured as well as the calculated intensity distribution exhibits a

strong dependence on the incident spin. An annular ring structure with a dark spot in the center for $|\sigma_+\rangle$ illumination and with a bright spot for $|\sigma_-\rangle$ illumination indicates coupling to different spiral plasmonic modes. The origin of the spin-dependent change in the near-field intensity distributions lies in the phase of the excited plasmonic mode. This phase was verified by FDTD calculation (see Fig. 1(g,h)) and was found to correspond to a spiral mode ($\exp(il\varphi)$, where $\varphi$ is the azimuthal angle) with topological charge $l = -2$ for $|\sigma_+\rangle$ and $l = 0$ for $|\sigma_-\rangle$ illumination.

To investigate the origin of the spin-dependent phase, we fabricated a simpler structure with full rotational symmetry (see Fig. 2(a)). This structure consisted of concentric rings with the same depth and periodicity as before and a central circular microcavity ($r_0 = 1.2\mu m$). The eigenmodes of such a circular plasmonic cavity are given by,

$$\mathbf{E}_{l,p}(\varphi, r) = E_0 \exp(ik_z z)[\exp(il\varphi)]J_l(k_p r)\hat{\mathbf{z}}, \quad (1)$$

where $k_z$ is the wavenumber perpendicular to the surface direction, $k_p$ is the radial wavenumber; $\varphi$, $r$ and $z$ are the cylindrical coordinate set, $l$ is the topological charge and $E_0$ is a constant. The notation $J_l$ stands for the $l$-order Bessel function of the first kind. The plasmonic in-plane wavenumber is given approximately by a delocalized plasmon relation, $k_p \sim k_0\sqrt{\varepsilon_2/(1+\varepsilon_2)}$, where $\varepsilon_2$ is the dielectric constant of the metal $(\text{Re}(\varepsilon_2) < -1)$ and $k_0 = 2\pi/\lambda_0$ is the wavenumber of the incident light. The wavenumber in the $z$ direction is expressed by $k_z^2 + k_p^2 = k_0^2$ and consequently is purely imaginary ($k_p > k_0$). Equation (1) represents an optical scalar vortex with a topological charge, *l*, which is associated with the optical angular momentum of the mode and corresponds to appropriate phase boundary conditions. The experimental near-field intensity distribution presented in Fig. 2(b) indicates a non-zero topological charge, which corresponds to a helical phase distribution for both $|\sigma_+\rangle$ and $|\sigma_-\rangle$ illumination. We verified this phase by the FDTD calculation presented in Fig. 2(d,e) and found that its helicity is given by $l = -\sigma_\pm$. This is evidence of a selective spin-dependent coupling to a single cavity mode induced by the structure. The existence of a spin-dependent spiral phase can be elucidated by analyzing the coupling mechanism between incident light with a specific spin and a surface plasmon cavity mode.

In plasmonic systems, when a coupling grating is illuminated by an arbitrarily polarized beam, the surface waves are excited via Transverse Magnetic (TM) polarization, corresponding to the magnetic field parallel to the grooves direction [15]. The propagation direction of the resulting plasmonic surface wave is perpendicular to the grating grooves whereas its polarization is linear in the vertical direction, $E_z$. In the proposed structure, the direction of the grating's grooves varies azimuthally, which results in the coupling of an incident planewave to a surface wave with a space-variant local direction. This non-trivial simultaneous manipulation in the space of polarizations and in the space of directions is most conveniently described using geometric representation upon the Majorana sphere [16,17]. In this model, polarized light is characterized by two dots on a unit sphere in the direction space (*x,y,z*), i.e., by two vectors, **u** and **v**, which point to these dots from the origin. Their bisector unit vector, **n**, coincides with the propagation direction of the wave, while its sign corresponds to the helicity of the polarization ellipse. For pure circular polarizations the bisector can be denoted as $\sigma_\pm$**n**. The projections of the dots onto the plane perpendicular to **n** denote the two foci of the corresponding polarization ellipse. The physical significance of this representation is that it provides a method to evaluate a geometric Berry phase [18,19] for complex fields whose polarization and direction have been modified.

Let us now consider right-handed circularly polarized illumination $|E_{ext}\rangle = E_0|\sigma_+\rangle$ impinging upon the element depicted in Fig. 2(a) from the bottom. Here $E_0$ is the amplitude of the field, which can be taken as one for brevity purposes. The circularly polarized incident beam, propagating along the *z* direction, can be depicted on the Majorana sphere by vectors **u** and **v**, which coincide at the north pole (see Fig. 3). The evanescent vertically polarized electric field that propagates on the metal surface in a radial direction is consequently represented by a vector **u`**, which still points to the north pole, and a vector **v`** which now points to the south pole. The interaction of light with a coupling grating whose local orientation is $\varphi$ can then be defined by a geodesic arc connecting the north pole with the south pole that intersects the *xy* plane at angle $\varphi$ (see Fig. 3). The Berry phase in our specific case is geometrically associated with the area enclosed on the Majorana sphere by the paths of vectors **u** and vector **v**, explicitly, $\phi_g = -(\Omega_\mathbf{u} + \Omega_\mathbf{v})/2$. Since the vector **u** is static, the Berry phase between the fields at two different azimuthal locations will be given simply as half of the area between two corresponding geodesics, which yields $\phi_g = -\varphi$. In the opposite case, when left-handed circularly polarized light is applied, i.e., $|E_{ext}\rangle = E_0|\sigma_-\rangle$, the picture on the sphere will be reversed. The initial state will be denoted by vectors **u** and **v** both pointing to the south pole, and the final state will be defined by **u`** pointing to the south pole and **v`** pointing to the north pole. In the second case, the path traversed on the Majorana sphere is in the reverse direction, which corresponds to the positive geometric phase, $\phi_g = \varphi$. The general case, therefore, can be given by $\phi_g = -\sigma_\pm \varphi$. Note that this phase results from the spin-orbit angular momentum coupling due to the space-variant polarization state and directional manipulations and is therefore geometric in nature. The appearance of the geometric phase in our experiment resulted from the SPP excitation by a spatially rotated grating and resembles the phase delay that arises when circularly polarized light is transmitted through a rotated polarizer [19]. This phase is linear with $\varphi$ and spirals around the center of the structure, giving rise to a phase singularity with a topological charge $l = -\sigma_\pm$; these properties explain the results presented in Fig. 2.

The intensity distributions in the cavity described above for incident spins $\sigma_+$ and $\sigma_-$ are indistinguishable, as opposed to the experimental results obtained in the spiral cavity (see Fig. 1). In the spiral structure, in addition to the geometric phase of the SPPs due to a polarization-dependent coupling, a dynamic phase arises as a result of a space-variant path difference. This dynamic phase is induced by the grooves' spiral pitch. The overall phase in the spiral cavity is the sum of the geometric and dynamic phases, $\phi = \phi_g + \phi_d$ and the total topological charge of the plasmonic vortex is $l = -(\sigma_\pm + m)$, where *m* is the spiral pitch in the units of SPP wavelength. For the specific case when *m*=1, the microcavity mode obeys the form of $J_0(k_p r)$ for $|\sigma_-\rangle$ and of $\exp(-i2\varphi)J_2(k_p r)$ for $|\sigma_+\rangle$. Consequently, for $|\sigma_-\rangle$ illumination the resultant field distribution possesses no phase singularity in the center, as opposed to the $|\sigma_+\rangle$ illumination case, which agrees with our experimental results presented in Fig. 1. The formation of the geometric phase is, therefore, the origin of the spin-dependent intensity distribution in our spiral cavity, and may lead to other unexpected spin-based phenomena in plasmonic systems.

One of the possible technological implementations of the plasmonic geometric phase could be a spin-dependent plasmonic focusing lens. We fabricated an appropriate structure that consisted of a 150nm-thick gold film with a hemi-circular coupling grating followed by a Bragg grating on the outer side (see Fig. 4(e)). The inner radius of the lens was $r_0 = 1.64$ μm. The structure was illuminated from the bottom with $|\sigma_+\rangle$ and $|\sigma_-\rangle$ planewaves and the intensity distribution was collected by the NSOM tip. The measured intensity of the plasmonic wave is presented in Fig 4(a,b). A spin-dependent transverse shift of the focus is easily observed by comparing the cross-sections of the spots (Fig. 4(c)). This shift can be regarded as a manifestation of the optical Magnus effect [20, 21] and the optical spin-Hall effect [22, 23] which arises in our system due to a spin-orbit coupling producing a spiral geometric phase. The experimental results were supported by the numerical simulation using FDTD which are also presented in the Fig. 4(c). The focal shift corresponds to the spin-dependent spiral phase modification due to the Berry phase and can be estimated by $\Delta x \approx r_0 |k_p^{-1}| |\nabla(\sigma_\pm \varphi) = \sigma_\pm / |k_p|$, ($\Delta x \approx 160 nm$). The measured shift of the focal intensity distribution is about 200 nm which is in good agreement with the above estimation as well as with the FDTD results (see Fig. 4(c)). The slight deviation of the measured results from the FDTD simulation can be attributed to the modified dispersion relation of the SPPs due to impurities caused by the fabrication process. This geometric effect is exceptionally pronounced when the superposition of spin $|\sigma_+\rangle$ and $|\sigma_-\rangle$ beams – linearly polarized light – is incident upon the structure. In this case, the focal spot is split in the lateral direction (see Fig. 4(d)). This experiment clearly reminds us of the effect of spin-dependent electron beam splitting in the Stern-Gerlach experiment [24, 25]. Thus, our structure can be considered as a unique polarization-dependent plasmonic element based on a geometric phase, which can be utilized for polarization demultiplexing, selective waveguide coupling and switching.

*Summary.* - We found a geometric phase of a surface-confined plasmonic wave that incorporates the interaction of light with an anisotropic inhomogeneous structure. A spiral plasmonic microcavity was demonstrated to produce a spin-dependent vortex mode with a spiral geometric phase induced by the spin-orbit interaction. A phenomenon of spin-dependent splitting of the focal spot of a plasmonic focusing lens was demonstrated, and explained in terms of a geometric phase. We have shown that in plasmonic systems, the phenomenon of the geometric phase cannot be ignored and, indeed, must be considered when dealing with the non-trivial interaction of light with surface waves. Further research is needed to investigate other spin-based effects and to propose a new generation of plasmonic geometric phase elements for various nano-photonic applications in the near-field.

**Figure Captions**

FIG. 1. Spin-dependent geometric phase in the spiral plasmonic structure. (a) The geometry of the structure (side view) and the optical setup. The Quarter Wave Plate (QWP) was used to generate a circularly polarized illumination, which was partially coupled to the SPP mode in the cavity – $E_z$. (b) Scanning electron microscope (SEM) picture of the spiral plasmonic cavity. (c,d) Intensity distribution in the cavity measured by a NSOM for $|\sigma_+\rangle$ and $|\sigma_-\rangle$ illumination, respectively. (e,f) Calculated intensity in the cavity for $|\sigma_+\rangle$ and $|\sigma_-\rangle$ illumination. (g,h) Calculated phase in the cavity for $|\sigma_+\rangle$ and $|\sigma_-\rangle$ illumination.

FIG. 2. Plasmonic field distribution in the circular cavity. (a) SEM picture of the circular structure. (b) Measured intensity distribution in the near-field for $|\sigma_+\rangle$ illumination. An identical intensity distribution was obtained for $|\sigma_-\rangle$. (c) Calculated intensity distribution for $|\sigma_+\rangle$. (d,e) Calculated $E_z$ field phase inside the cavity for $|\sigma_+\rangle$ and $|\sigma_-\rangle$ illumination, respectively.

FIG. 3. Graphical representation of the geometric phase on the Majorana sphere. Geometric phase induced by the interaction of light with an anisotropic inhomogeneous structure is equal to half of the area enclosed by the two geodesics that correspond to different propagation directions, $k_p$ of the excited SPPs (gray area). The variation of the propagation wavevector $\mathbf{k}_p(\varphi)$ along the element is depicted in the small picture by blue arrows.

FIG. 4. Spin-dependent plasmonic lens based on a geometric phase. (a, b) The intensity distributions measured by a NSOM for $|\sigma_+\rangle$, and $|\sigma_-\rangle$ illumination, respectively. (c) The transverse cross-sections of the measured intensity distributions in the focal plane of the lens for $|\sigma_+\rangle$ (blue squares) , $|\sigma_-\rangle$ (red circles) illumination, respectively. FDTD calculation is plotted for each polarization (solid blue line - $|\sigma_+\rangle$; dashed red line - $|\sigma_-\rangle$). The cross-sections in (c) were measured along the horizontal dashed lines depicted in (a) and (b). (d) The intensity distribution measured by NSOM for $|\sigma_+\rangle + |\sigma_-\rangle$ (linearly polarized) illumination. (e) The SEM picture of the element.

**Figures**

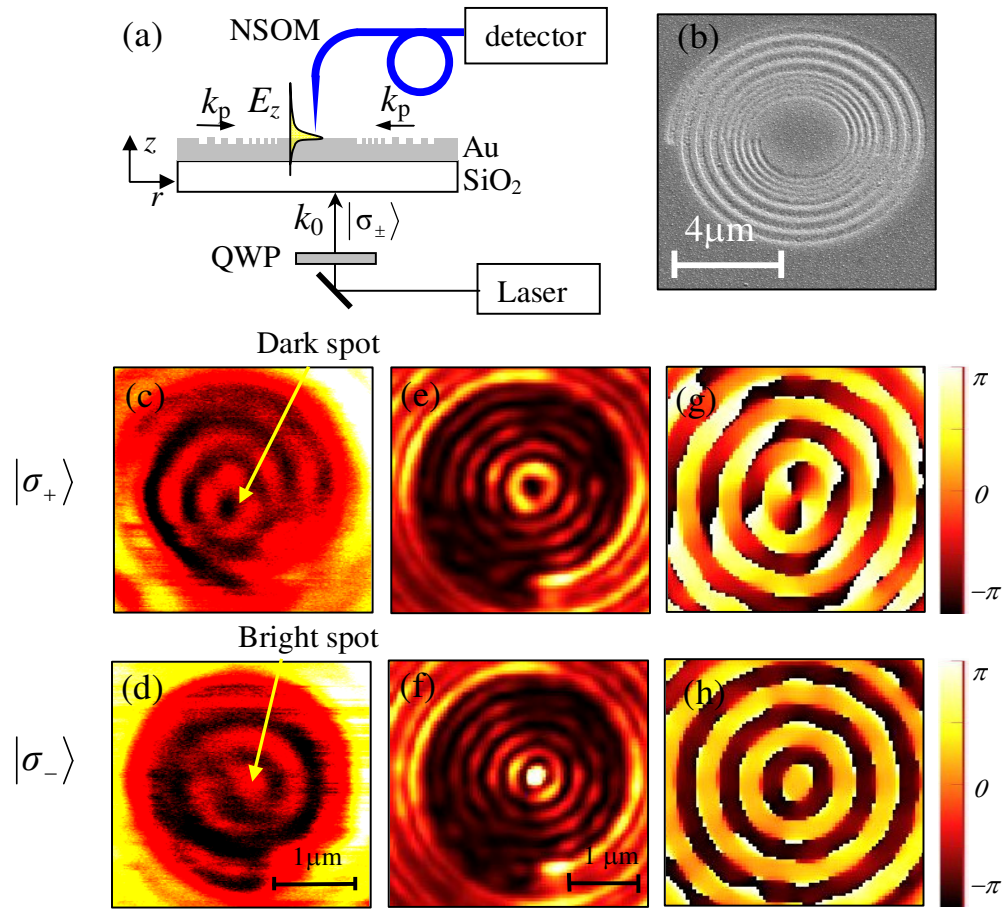

FIG. 1. Y. Gorodetski et al.

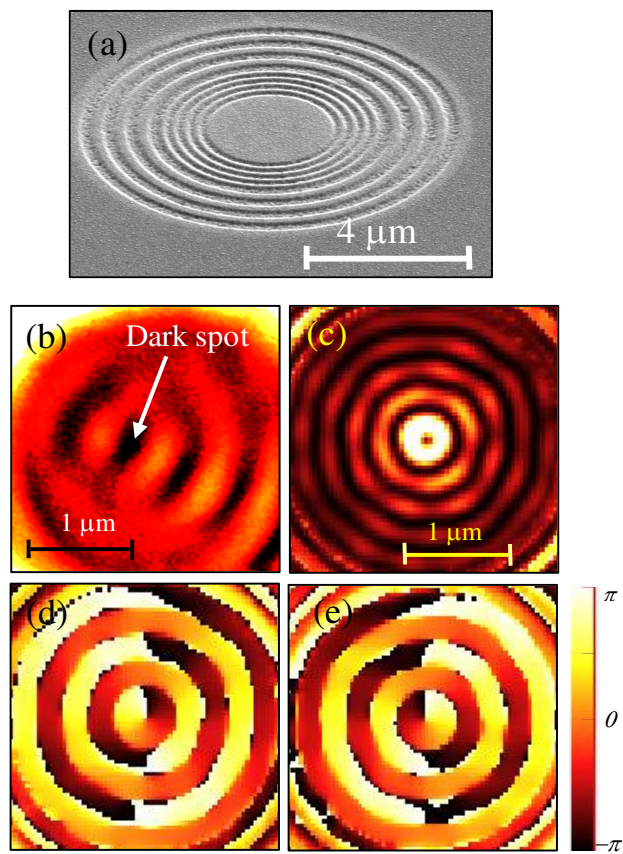

FIG. 2. Y. Gorodetski et al.

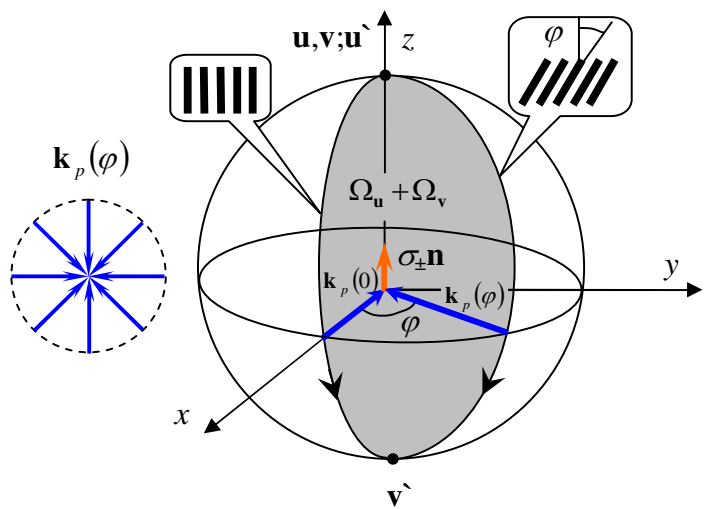

FIG. 3. Y. Gorodetski et al.

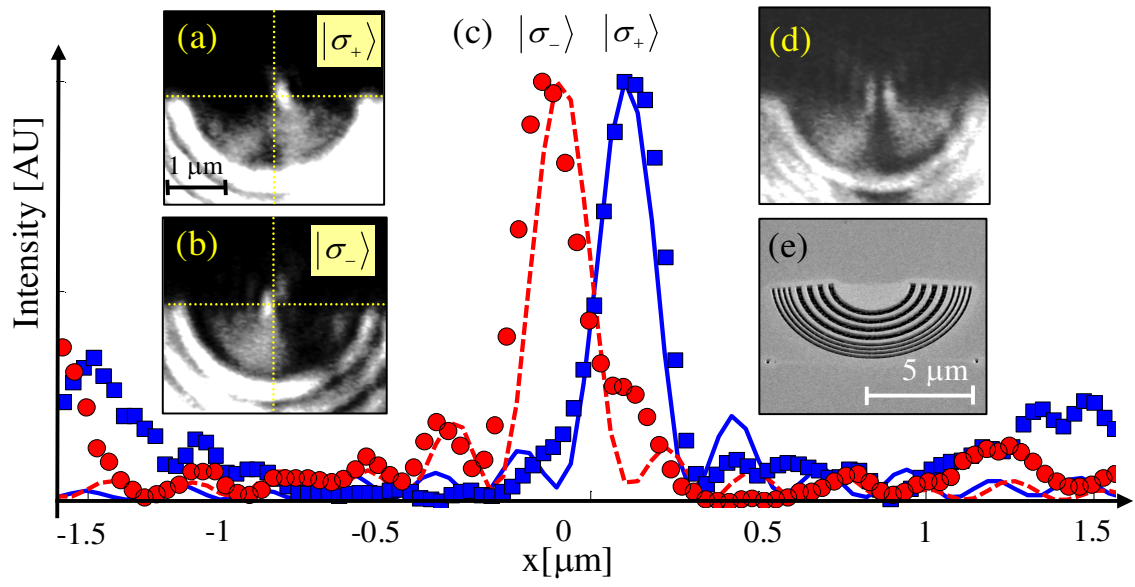

FIG. 4. Y. Gorodetski et al.